%% file: paper.tex
\documentclass{article}
\usepackage{spconf,amsmath,amssymb,amsthm,epsfig,color,empheq,graphicx,graphics,balance,algorithm,algorithmic,braket}
\usepackage{url,wasysym,epstopdf,enumitem,array}
\usepackage{multirow}
\usepackage{xcolor}

\title{A QUANTUM APPROACH FOR STOCHASTIC CONSTRAINED BINARY OPTIMIZATION}
\name{Sarthak Gupta and Vassilis Kekatos\thanks{This work was supported by a seed funding grant from the Virginia Commonwealth Cybersecurity Initiative (CCI) -- Southwest Virginia node.}}
\address{Bradley Dept. of ECE, Virginia Tech, Blacksburg, VA 24061, USA; \{gsarthak,kekatos\}@vt.edu}
\input{stylesV2}

\begin{document}
\maketitle

\begin{abstract}
Analytical and practical evidence indicates the advantage of quantum computing solutions over classical alternatives. Quantum-based heuristics relying on the variational quantum eigensolver (VQE) and the quantum approximate optimization algorithm (QAOA) have been shown numerically to generate high-quality solutions to hard combinatorial problems, yet incorporating constraints to such problems has been elusive. To this end, this work puts forth a quantum heuristic to cope with stochastic binary quadratically constrained quadratic programs (QCQP). Identifying the strength of quantum circuits to efficiently generate samples from probability distributions that are otherwise hard to sample from, the variational quantum circuit is trained to generate binary-valued vectors to approximately solve the aforesaid stochastic program. The method builds upon dual decomposition and entails solving a sequence of judiciously modified standard VQE tasks. Tests on several synthetic problem instances using a quantum simulator corroborate the near-optimality and feasibility of the method, and its potential to generate feasible solutions for the deterministic QCQP too. 
\end{abstract}

\begin{keywords}
QAOA, VQE, dual decomposition, quantum unconstrained binary optimization (QUBO). 
\end{keywords}

\section{INTRODUCTION}\label{sec:intro}
Quantum computers exhibit an innate ability to handle exponentially large computations in a parallel fashion yet with a strong probabilistic flavor. Quantum algorithms such as Shor's integer factorization, Grover's search, and the linear system solver of Harrow-Hassidim-Lloyd (HHL) can attain polynomial or even exponential speedups over the best known algorithms on classical computers~\cite{nielsen00}. Nonetheless, some of these quantum algorithms presume a large number of qubits on fault-tolerant quantum computers. In the \emph{near-term intermediate scale} (NISQ) era, quantum computers are noisy and thus oftentimes limited in terms of number of gates and/or qubits. With such limitations in mind, \emph{variational} quantum algorithms have been suggested as promising tools to practically showcase quantum advantage in the NISQ setup~\cite{Simeone}.

Variational quantum computers involve a sequence of parameterized gates. Their parameters are updated externally by a classical computer in a closed-loop fashion to steer the quantum state towards a desirable direction. The variational quantum eigensolver (VQE) used to provide high-quality solutions to combinatorial problems is a representative example. The Quantum Approximate Optimization Algorithm (QAOA)~\cite{Farhi14} is a special instance of VQE. In QAOA, not only the parameters but also the architecture of the quantum circuit become problem-dependent. The quantum circuit trained by QAOA operates as a sampler to efficiently generate near-optimal solutions of binary quadratic problems (e.g., MAXCUT); see~\cite{Hadfield19} for a summary of claims on QAOA. 

While most VQE/QAOA schemes target unconstrained problems, dealing with constraints is crucial to several applications in machine learning, wireless communications, and financial (stock trading) optimization. Adding constraints to QAOA or adiabetic quantum computing~\cite{McGeoch} (the QAOA counterpart for non-gate-based quantum computers) has been pursued in two ways. One approach has been to convert the constrained problem into an unconstrained minimization of a Lagrangian-like function~\cite{Lucas,Ohzeki}. However, the weights for constraint penalties can be safely selected only if constraints are expressed as Boolean functions or linear equalities. An alternative approach modifies the architecture of the quantum circuit (via the mixer Hamiltonian of QAOA) to confine quantum states on the subspace spanned by constraints~\cite{HenSarandy,HenSpedalieri,Hadfield19,Hadfield17}. Nonetheless, constructing such `driver' mixer Hamiltonians is again highly problem-dependent and often limited to equality constraints. Reference~\cite{Ronagh} develops a quantum adiabetic approach to tackle binary linearly-constrained quadratic programs. It targets the dual problem and interfaces the quantum computer with a branch-and-bound scheme ran classically. Reference~\cite{GambellaSimonetto} treats mixed-binary quadratic-plus-convex problems using the alternating direction method of multipliers (ADMM) to split binary and continuous variables into separate minimizations, solved by QAOA and classical convex optimizers respectively per ADMM iteration.

\emph{Relation to prior work.} Addressing binary QCQPs by quantum heuristics has been largely unexplored to the authors' knowledge. We put forth a quantum-based heuristic to solve a stochastic binary QCQP. Harnessing the power of quantum circuits to sample from probability mass functions (PMF) that are hard to sample classically, we devise a dual decomposition technique that solves a sequence of standard VQE tasks to systematically adjust Lagrangian multipliers. Numerical tests using quantum computer simulators provided by IBM evaluate this technique on randomly generated stochastic and deterministic binary QCQPs. 


\vspace*{-1em}

\section{Quantum Computing Preliminaries}\label{sec:qc}
A quantum system consisting of $n$ quantum bits (qubits) is described by an exponentially large state vector $\ket{\bx}\in\mathbb{C}^N$ with $N=2^n$ assuming the system is in a \emph{pure state}. The Dirac notation $\ket{\bx}$ named \emph{ket} emphasizes that vector $\bx$ is unit-norm or $\sum_{k=0}^{N-1}|x_k|^2=1$. If $\be_k$ is the $k$-th canonical vector of length $N$, we can write $\ket{\bx}=\sum_{k=0}^{N-1}x_k\ket{\be_k}$. The vector $\be_k$ is oftentimes alternatively expressed as $\ket{\be_k}=\ket{k}$, where $k$ is the binary representation of index $k$. For example, a system with $n=2$ qubits has a state in $\mathbb{C}^4$, which is spanned by canonical vectors $\{\be_k\}_{k=0}^3$ and $\be_0=[1~0~0~0]^\top=\ket{00}$. Vector $\ket{\bx}$ provides a statistical characterization for the quantum state: If we measure the quantum system output, its qubits will be in configuration $\ket{k}$ with probability $|x_k|^2$ for all $k$. Symbol $\bra{\bx}$ termed \emph{bra} denotes the conjugate transpose of $\ket{\bx}$, while the \emph{braket} $\braket{\bx | \by}$ denotes the inner product between states. 

The fundamental operations we can perform on a quantum system is evolution and measurement. The former can be described by the application of a unitary $\bU$ on state $\ket{\bx}$ to produce state $\ket{\by}=\bU\ket{\bx}$. Although $\bU$ is exponentially large, it is usually implemented efficiently using quantum gates. Among various types of measurements, we focus on \emph{projective measurements}. A projective measurement is associated with a Hermitian matrix (named \emph{observable}) and its eigenvalue decomposition $\bH=\sum_{m=1}^M\lambda_m\bv_m\bv_m^H$. If such measurement is performed on $\ket{\bx}$, outcome $m$ is observed with probability $p_m:=|\braket{\bx | \bv_m}|^2$. Define a random variable taking value $\lambda_m$ when outcome $m$ is observed. The expected value of this variable is $\braket{\bx | \bH | \bx}=\sum_{m=1}^M p_m \lambda_m$. If $\bH$ is diagonal, the measurement is on the \emph{computational basis}. This is practically important because now $\bv_m=\be_m$, outcome $m$ relates to $\ket{m}$, and each qubit can be measured individually.

If quantum system $i$ has been prepared in state $\ket{\bx_i}$ for $i=1,2$, their joint state would be $\ket{\bx_1}\otimes\ket{\bx_2}$, where $\otimes$ is the Kronecker product. This is oftentimes represented as $\ket{\bx_1}\ket{\bx_2}$ or $\ket{\bx_1,\bx_2}$. The Kronecker product rule generalizes to the composition of $n$ systems. For example, $\ket{1}\ket{1}\ket{0}=\be_1\otimes \be_1\otimes \be_0=\be_6=\ket{110}$, where the canonical vectors shown in the middle are in $\mathbb{R}^2$ and those at the end are in $\mathbb{R}^8$.

\section{Variational Quantum Eigensolver (VQE)}\label{sec:vqe}
VQE is a heuristic approach to find near-optimal solutions for combinatorial problems of the general form
\begin{equation}\label{eq:problem}
    \min_{\bb\in\{0,1\}^n} f(\bb).
\end{equation}
A particular example of interest is the quadratic unconstrained binary optimization (QUBO) problem with
\begin{equation}\label{eq:qubo}
    f(\bb)=\bb^\top\bA\bb +\bb^\top \bc+d
\end{equation}
which is known to be NP-hard. For later developments, it is convenient to reformulate QUBO in terms of the \emph{spin} $\{\pm1\}$ variables through the transformation
\begin{equation}\label{eq:trans}
s_i=1-2b_i=(-1)^{b_i}~~\text{for}~~i=0,\ldots,n-1.
\end{equation}
Collecting the spin variables in vector $\bs=\bone-2\bb$, the quadratic objective can be equivalently expressed as
\begin{equation}\label{eq:qubo2}
    f(\bb)=\bar{f}(\bs)=\bs^\top\bbA\bs +\bs^\top \bbc+\bar{d}
\end{equation}
where $\bbA:=\tfrac{1}{4}\bA$; $\bbc:=-\frac{1}{2}(\bA\bone+\bc)$; and $\bar{d}:=\tfrac{1}{4}\bone^\top\bA\bone+\tfrac{1}{2}\bone^\top\bc+d$. We next explain how VQE samples high-quality solutions of~\eqref{eq:problem} by solving an eigenvalue minimization task.

The VQE method falls under the family of \emph{variational} quantum algorithms. The term \emph{variational} pertains to the fact that the quantum circuit is not fixed, but parameterized by relatively few parameters collected in vector $\btheta\in\mathbb{R}^P$. These parameters are iteratively adjusted by classical computer in a closed-loop fashion so that the quantum system eventually reaches a desirable state. The process resembles the training of a neural network whose weights are updated by an optimization algorithm. Similarly to neural networks where the learner has to select an architecture (e.g., network depth/width and type of activations), the parameterized form (also termed \emph{ansatz}) of the variational quantum circuit is specified \emph{a priori}. We will be using a 2-local ansatz where single-qubit $R_Y$ gates are applied to all qubits, followed by a full entanglement circuit, all repeated for 3 layers (iterations)~\cite{Simeone}.

Given $\btheta$ and driven by input $\ket{0}^n$, the quantum circuit produces at its output the quantum state $\ket{\bx(\btheta)}=\bU(\btheta)\ket{0}^n$ for a unitary $N\times N$ matrix $\bU(\btheta)$. To simplify notation, we will oftentimes write $\ket{\bx}$ in lieu of $\ket{\bx(\btheta)}$. Albeit $\ket{\bx}\in \mathbb{C}^N$ is exponentially long, it can be easily generated by the quantum circuit though it cannot be read out of the circuit as a vector in a computationally efficient manner. Instead, it is relatively easy to sample from it. Every time we run the quantum circuit driven by $\ket{0}^n$, we will be observing one of the binary outputs $\ket{k}=\ket{\be_k}$ with probability $p_k:=|x_k|^2$ for $k=0,\ldots,N-1$. The quantum circuit thus serves as an efficient sampler from the exponentially large probability mass function (PMF) $\{p_k\}_{k=0}^{N-1}$. 

To exploit this sampling property, we next relate the cost $f(\bb)$ with a so-termed \emph{Hamiltonian} matrix $\bH$ so that 
\begin{equation}\label{eq:eigenproperty}
    \bH\ket{\be_k}=f(\ket{k})\ket{\be_k}\quad \text{for all}~k.
\end{equation}
Matrix $\bH$ is apparently diagonal and carries all $N$ function evaluations $f(\be_k)$ on its diagonal. Moreover, the canonical vectors $\be_k$ constitute the eigenvectors of $\bH$, each with corresponding eigenvalue $f(\ket{k})$. Therefore, the minimization in~\eqref{eq:problem} can be reformulated as the problem of finding the eigenvector corresponding to the minimum eigenvalue of $\bH$
\begin{equation}\label{eq:eigenproblem}
    \min_{\ket{\bx}} \bra{\bx}\bH\ket{\bx}.
\end{equation}
As long as $\ket{\bx}$ is allowed to take any of the values $\{\be_k\}_{k=0}^{N-1}$, the minimizer of \eqref{eq:eigenproblem} corresponds to the minimizer of \eqref{eq:problem}. For example, if a quantum system has $n=3$ qubits, its state would be $\ket{\bx}\in\mathbb{C}^8$. Here $\be_k$'s are the columns of the identity matrix $\bI_8$. If the minimizer of \eqref{eq:eigenproblem} is $\ket{\be_5}=\ket{b_1b_2b_3}=\ket{101}$, then the minimizer of \eqref{eq:problem} is $\bb=[1~0~1]^\top$; and vice versa. 

Although $\bH$ is exponentially large, it can be implemented using only $\mcO(n^2)$ quantum gates since it can be expressed as
\begin{equation}\label{eq:H}
    \bH=\sum_{i=0}^{n-1}\sum_{j=0}^{n-1}\bar{A}_{ij}\bZ_i\bZ_j 
    +\sum_{i=0}^{n-1}\bar{c}_{i}\bZ_i
    +\bar{d}\bI_N
\end{equation}
where the $N\times N$ Hermitian matrix $\bZ_i$ is defined as 
\begin{equation*}
    \bZ_i=\bI_2\otimes\cdots\otimes\bZ\otimes\cdots\otimes\bI_2~~\text{with}~~
    \bZ=\left[\begin{array}{cc}
    1 &0\\
    0 &-1
    \end{array}\right].
\end{equation*}
This is a Kronecker product involving $(n-1)$ identity matrices $\bI_2$ and one \emph{Pauli-Z} operator $\bZ$ applied to the $i$-th qubit. Matrix $\bH$ as defined in \eqref{eq:H} is obviously diagonal. To establish \eqref{eq:eigenproperty}, note first that $\bZ\ket{0}=\ket{0}$ and $\bZ\ket{1}=-\ket{1}$, or more compactly, $\bZ\ket{b}=(-1)^b\ket{b}$. Consequently, when $\bZ_i$ is applied to a state $\ket{\bb}=\ket{b_1b_2\cdots b_n}$, the effect is $\bZ_i\ket{\bb}=(-1)^{b_i}\ket{\bb}=s_i\ket{\bb}$ from \eqref{eq:trans}. Similarly, it also holds that $\bZ_i\bZ_j\ket{\bb}=s_is_j\ket{\bb}$. Property~\eqref{eq:eigenproperty} now follows immediately by postmultiplying \eqref{eq:H} by any $\ket{\be_k}$ and using $f(\bb)=\bar{f}(\bs)$.

If $\ket{\bx}$ in \eqref{eq:eigenproblem} is restricted to set $\mcE:=\{\be_k\}_{k=0}^{N-1}$, problem \eqref{eq:eigenproblem} is as hard as \eqref{eq:problem}. VQE relaxes \eqref{eq:eigenproblem} to the set of all quantum states $\ket{\bx(\btheta)}$ that can be parameterized by the chosen ansatz and via $\btheta$. Problem~\eqref{eq:eigenproblem} is then solved over $\btheta$ rather than $\ket{\bx}$
\begin{equation}\label{eq:eigenproblem2}
    \min_{\btheta}~F(\btheta):=\braket{\bx(\btheta) | \bH | \bx(\btheta)}.
\end{equation}

From the eigenvalue property \eqref{eq:eigenproperty}, it follows $\bra{\be_n}\bH\ket{\be_k}=f(\ket{k})$ for all $k$. How about $\bra{\bx}\bH\ket{\bx}$ for a general state $\ket{\bx}$? Because $\ket{\bx}=\sum_{k=0}^{N-1}x_k\ket{\be_k}$, it is easy to show that
\begin{equation}
    \braket{\bx | \bH | \bx}=\sum_{k=0}^{N-1}|x_k|^2 f(\ket{k})=\sum_{k=0}^{N-1}p_k f(\ket{k}).
\end{equation}
In other words, function $F(\btheta)$ is the average of $f$ under the PMF defined by $\ket{\bx}$. For instance, the random outcome $\ket{k}=\ket{101}$ occurring with probability $|x_5|^2$ is assigned to the random variable $f$ taking the value $f([1~0~1]^\top)$. Hence, function $F(\btheta)$ is really an expectation (an \emph{observable} in the quantum computation parlance) of function $f(\bb)$ when $\bb$ is drawn from the PMF $\{|x_k(\theta)|^2\}_{k=0}^{N-1}$. Ideally, the global minimizer $\btheta$ of \eqref{eq:eigenproblem2} defines a PMF via $\ket{\bx(\btheta)}$ that samples with non-zero probability only the canonical vectors $\ket{\be_k}$ associated with the smallest eigenvalue of $\bH$.


Problem~\eqref{eq:eigenproblem2} is solved in a hybrid fashion: The quantum computer samples from $\ket{\bx(\btheta)}$ and estimates $F(\btheta)$ and possibly its gradient $\nabla_{\btheta} F$. A classical computer uses the previous information and iteratively updates $\btheta$ based on a zero- or first-order optimization algorithm, such as gradient descent or Bayesian optimization. As with training neural networks, $F(\btheta)$ is nonconvex due to the form of the ansatz. Moreover, the ensemble statistic $F(\btheta)$ cannot be computed exactly, but estimated as the sample average $\hat{F}(\btheta):=\sum_{r=1}^R f(\bb_r)/R$ over $R$ runs, where $\bb_r$ is the quantum output after run $r$. 


\section{CONSTRAINED VQE}\label{sec:cVQE}
As discussed earlier, VQE provides a successful heuristic for solving QUBO through the variational formulation of \eqref{eq:eigenproblem2}. Can VQE be generalized to deal with a binary QCQP of the ensuing form?
\begin{align}\label{eq:cproblem}
\min_{\bb\in\{0,1\}^n}~&~f_0(\bb)\\ 
\mathrm{s.to}~&~f_m(\bb)\leq 0,\quad m=1:M.\nonumber
\end{align}
Here $f_m(\bb):=\bb^\top\bA_m\bb+\bb^\top\bc_m+d_m$ for $m=0,\ldots,M$. Solving such problems is also known to be NP-hard. Providing a quantum heuristic to directly deal with \eqref{eq:cproblem} seems to be challenging. To this end, we relax expectations and aim at designing a quantum state $\ket{\bx}$ from which we can draw binary-valued $\bb$ that solve the \emph{stochastic} binary QCQP:
\begin{align}\label{eq:sampling}
\min_{\ket{\bx}}~&~\mathbb{E}_{\bx}[f_0(\bb)]\\ 
\mathrm{s.to}~&~\mathbb{E}_{\bx}[f_m(\bb)]\leq 0,\quad m=1:M.\nonumber
\end{align}
As in the unconstrained setup, rather than minimizing over $\ket{\bx}$, we propose optimizing over a PMF parameterized by $\btheta$ and captured by quantum state $\ket{\bx(\btheta)}$. Specifically, we suggest solving the constrained minimization
\begin{align}\label{eq:ceigenproblem}
\min_{\btheta}~&~F_0(\btheta)\\ 
\mathrm{s.to}~&~F_m(\btheta)\leq 0:\quad \lambda_m,\quad m=1:M\nonumber
\end{align}
where each observable $F_m(\btheta):=\braket{\bx(\btheta)|\bH_m |\bx(\btheta)}$ depends on the Hamiltonian $\bH_m$ defined similar to $\bH$ in \eqref{eq:H} for all $m$. Heed that problem~\eqref{eq:ceigenproblem} can be reformulated and solved as a linear program (LP) over the PMF of $\bb$. Nonetheless, that requires evaluating $\{f_m(\bb)\}_{m=0}^M$ for all $2^n$ values of $\bb$. Moreover, the optimization variable of this LP is the vector of PMF values that is exponentially large too. That is also the case with standard VQE/QAOA.

Contrary to \eqref{eq:cproblem}, problem \eqref{eq:ceigenproblem} is over the continuous variable $\btheta$, and thus, we can associate a dual variable $\lambda_m$ for each constraint and define its Lagrangian function
\begin{equation}\label{eq:Lagrangian}
    L(\btheta;\blambda):=F_0(\btheta)+\sum_{m=1}^M \lambda_m F_m(\btheta)
\end{equation}
where $\blambda\in\mathbb{R}^M$ collects all dual variables. Problem \eqref{eq:ceigenproblem} could be solved via \emph{dual decomposition}, according to which $\blambda$ is updated iteratively via a subgradient ascent step on $L$ as
\begin{equation}\label{eq:du}
    \lambda_m^{t+1}:=\max\left\{\lambda_m^t+\mu_t F_m(\btheta^t),0\right\},~~ m=1:M 
\end{equation}
for a positive step size $\mu_t=\mu_0/(t+\alpha)$ with $\alpha>0$, and $\btheta^t$ is a minimizer of the Lagrangian $L(\btheta;\blambda^t)$ evaluated at $\blambda^t$:
\begin{equation}\label{eq:op}
    \btheta^t\in \arg\min_{\btheta} \braket{\bx(\btheta) |\bH_0+\sum_{m=1}^M\lambda_m^t \bH_m|\bx(\btheta)}.
\end{equation}
Problem~\eqref{eq:op} takes the QUBO form of \eqref{eq:eigenproblem2}, and is therefore amenable to standard VQE or even the celebrated QAOA approach. Under the latter, the ansatz takes a particular form that depends on the problem Hamiltonian $\bH_0+\sum_{m=1}^M\lambda_m^t \bH_m$. Here, we used a problem-independent ansatz under the general VQE framework and leave QAOA for future work. 

\section{NUMERICAL TESTS}\label{sec:tests}
The novel solver for~\eqref{eq:ceigenproblem} was implemented in Python using the Qiskit library~\cite{Qiskit}. The \emph{VQE} class in Qiskit was used to solve the minimization for the primal update~\eqref{eq:op}. In addition to providing the ansatz described in Section~\ref{sec:vqe}, the VQE class was configured with the `SLSQP' optimizer. The  maximum number of iterations was set to $1,000$, and we used the \texttt{aer\_simulator\_statevector} quantum simulation backend. For the dual update in~\eqref{eq:du}, constraint violations were measured over the observables $\bH_m$ using the minimum eigenstate returned by VQE. The stopping criteria $\|\blambda^{t}-\blambda^{t-1}\|_2\leq1\cdot10^{-5}$ was utilized to ascertain the convergence of the dual updates~\eqref{eq:du}.

To illustrate the application of the proposed strategy to solving the stochastic binary QCQP in~\eqref{eq:sampling}, several $2$-bit problem instances were sampled randomly by drawing the entries of $\{\bA_0,\bc_0,\bd_0\}$ and $\{\bA_1,\bc_1,\bd_1\}$ from the standard normal distribution, while ensuring the resulting problem was feasible. The VQE approach was compared against a linear program that finds a PMF solving~\eqref{eq:ceigenproblem}; this was possible due to the small value of $2^n$. For the two approaches, the obtained PMFs along with the associated dual variables are reported in Table~1 for 4 randomly sampled problem instances. 

\begin{table}[t]\label{mytable}
\centering
\caption{Comparing the exact solution of~\eqref{eq:ceigenproblem} obtained via a linear program and the proposed quantum-based approach.}
\vspace*{1em}
\begin{tabular}{|r|rr|rr|}
\hline\hline
\multicolumn{1}{|c|}{\multirow{2}{*}
{{\#}}} & \multicolumn{2}{c|}{\textbf{Found PMF}}                                      & \multicolumn{2}{c|}{\textbf{Dual}}                                    \\ \cline{2-5} 
\multicolumn{1}{|c|}{}                              & \multicolumn{1}{c|}{{Quantum}}           & \multicolumn{1}{c|}{{LP}} & \multicolumn{1}{c|}{{Quantum}} & \multicolumn{1}{c|}{{LP}} \\ \hline\hline
{1}                                          & \multicolumn{1}{l|}{$[0.44, 0, 0.56,  0]$} & {$[0.44, 0, 0.56, 0]$}           & \multicolumn{1}{l|}{$0.854$}         & $0.851 $                           \\ \hline
{2}                                          & \multicolumn{1}{l|}{$[0.71, 0, 0.29, 0]$}  & {$[0.70, 0, 0.30, 0]$}           & \multicolumn{1}{l|}{$0.337$}         & $0.337$                            \\ \hline
{3}                                          & \multicolumn{1}{l|}{$[0, 0.80, 0, 0.20]$}  & {$[0, 0.80, 0, 0.20]$}           & \multicolumn{1}{l|}{$0.459$}         & $0.459$                            \\ \hline
{4}                                          & \multicolumn{1}{l|}{$[0, 0, 0.61, 0.39]$}  & {$[0, 0, 0.60, 0.40]$}           & \multicolumn{1}{l|}{$0.566$}         & $0.566$                            \\ 
\hline\hline
\end{tabular}
\end{table}

\begin{figure}[t]
\centering
	\includegraphics[scale=0.5]{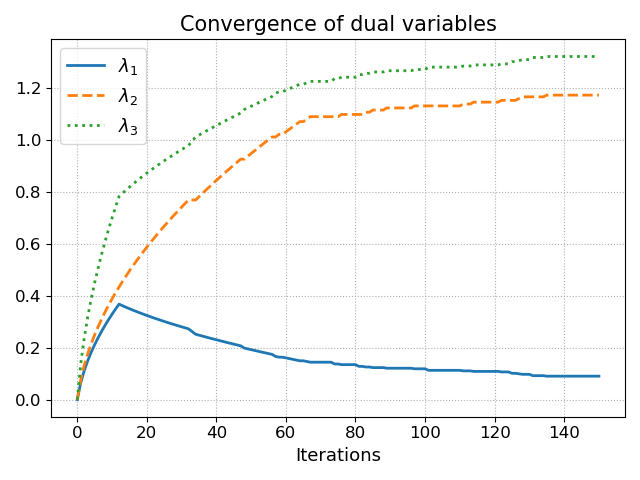}
	\caption{Convergence of dual variables under dual updates~\eqref{eq:du} for a stochastic binary QCQP with $M=3$ constraints.}
	\label{fig:dual_conv}
\end{figure}

To study the scalability of the approach and to verify the compatibility of the solutions with the deterministic QCQP in~\eqref{eq:cproblem}, we also sampled $30$ feasible 5-bit problem instances with three constraints each. The quadratic cost and constraint functions were generated as in the previous test. To avoid instances with non-binding constraints, the constants $d_m$ in the constraint functions were manually adjusted so that at least one of the constraints was active and yielded a non-zero dual variable. From the sampled problems, it was found that the dual decomposition involving VQE was able to produce the optimal solutions for 28 out of the 30 problem instances tested, whereas infeasible binary candidates were obtained for the remaining 2 instances. Figure~\ref{fig:dual_conv} illustrates the convergence of the dual variables for one of the problem instances, where all three constraints were found to be active.

\section{CONCLUSIONS}\label{sec:conclusions}
A novel generalization of VQE to address the need for dealing with stochastic binary QCQPs has been developed. Leveraging dual decomposition, the approach entails solving a sequence of judiciously modified VQE tasks. Numerical tests demonstrate that upon convergence of the constrained VQE algorithm, the variational quantum circuit is able to sample from a stochastic policy to generate binary-valued vectors that minimize the binary QCQP and satisfy its constraints in expectation. Some of these samples seem to be feasible for the deterministic binary QCQP too. This novel heuristic sets the foundation for further developments towards constrained discrete optimization. We are currently exploring several exciting directions: \emph{i)} Coupling this approach with QAOA rather than VQE; \emph{ii)} skipping the nested optimization in \eqref{eq:op} through a primal-dual decomposition alternative as in~\cite{GMDK21,OPFandLearn}; and \emph{iii)} dealing with mixed-binary setups.

\balance

\vfill\pagebreak

\bibliographystyle{IEEEbib}
\bibliography{quantum,kekatos}
\end{document}

%% file: stylesV2.tex
\newcommand \bone{\mathbf{1}}

\newcommand \bb{\mathbf{b}}
\newcommand \bc{\mathbf{c}}
\newcommand \bd{\mathbf{d}}
\newcommand \be{\mathbf{e}}

\newcommand \bs{\mathbf{s}}

\newcommand \bv{\mathbf{v}}

\newcommand \bx{\mathbf{x}}
\newcommand \by{\mathbf{y}}

\newcommand \bA{\mathbf{A}}

\newcommand \bH{\mathbf{H}}
\newcommand \bI{\mathbf{I}}

\newcommand \bU{\mathbf{U}}

\newcommand \bZ{\mathbf{Z}}

\newcommand \btheta{\boldsymbol{\theta}}

\newcommand \blambda{\boldsymbol{\lambda}}

\newcommand \mcE{\mathcal{E}}

\newcommand \mcO{\mathcal{O}}

\newcommand \bbc{\bar{\mathbf{c}}}

\newcommand \bbA{\bar{\mathbf{A}}}

